\documentclass[manuscript]{acmart}

\copyrightyear{2025}
\setcopyright{rightsretained}
\acmDOI{10.1145/3706599.3706687}

\begin{document}

\title{UX Challenges in Implementing an Interactive B2B Customer Segmentation Tool}

\author{Muhammad Raees}
\email{mraees@mail.rit.edu}
\orcid{0000-0002-1581-0378}
\affiliation{%
  \institution{Rochester Institute of Technology}
  \city{Rochester}
  \state{New York}
  \country{USA}
}

\author{Vassilis-Javed Khan}
\email{javed.khan@sappi.com}
\orcid{0000-0002-7333-981X}
\affiliation{%
  \institution{Sappi, Europe}
  \city{}
  \country{Belgium}
}

\author{Konstantinos Papangelis}
\email{kxpigm@rit.edu}
\orcid{0000-0001-5094-9837}
\affiliation{%
 \institution{Rochester Institute of Technology}
 \city{Rochester}
 \state{New York}
 \country{USA}}

\renewcommand{\shortauthors}{Raees et al.}

\begin{abstract}
In our effort to implement an interactive customer segmentation tool for a global manufacturing company, we identified user experience (UX) challenges with technical implications. 
The main challenge relates to domain users' effort, in our case sales experts, to interpret the clusters produced by an unsupervised Machine Learning (ML) algorithm, for creating a customer segmentation. 
An additional challenge is what sort of interactions should such a tool support to enable meaningful interpretations of the output of clustering models.
In this case study, we describe what we learned from implementing an Interactive Machine Learning (IML) prototype to address such UX challenges. 
We leverage a multi-year real-world dataset and domain experts' feedback from a global manufacturing company to evaluate our tool. 
We report what we found to be effective and wish to inform designers of IML systems in the context of customer segmentation and other related unsupervised ML tools.
\end{abstract}



\maketitle
\section{Introduction}
Customer segmentation is an established business method to improve data-driven decision-making based on grouping customers into segments to personalize treatment~\cite{manjunath2021distributed, cortez2021b2b}.
Segmentation methods help analyze large, high-dimensional datasets, such as sales transactions, to extract insights and decision indicators~\cite{paschen2020collaborative, iyelolu2024implementing}.
Among segmentation methods, unsupervised learning, particularly clustering, is often used to extract patterns and group (or differentiate) data into related segments without predefined outputs~\cite{xu2015comprehensive, bae2020interactive, cortez2021b2b}.
Unsupervised learning involves training a model on unlabeled data, which lacks predefined labels or categories for output~\cite{xu2015comprehensive}.
These methods are useful to extract meaningful insights, allowing businesses to make sense of unlabeled data~\cite{naim2023consumer, moradi2022applications}.

However, making sense of these unsupervised results can be challenging~\cite{mozolewski2022explain, raees2023four}.
For instance, clustering methods often generate semantically agnostic segments, requiring expert evaluation for meaningful interpretation~\cite{bae2020interactive}.
Interpreting clustering results is also an intrinsically subjective task correlated with domain knowledge, requiring users to examine model outputs and provide input/feedback to make meaningful understanding~\cite{mozolewski2022explain, bae2020interactive, bobek2023knac}.
This creates a two-fold problem; 1) semantic agnosticism makes it effortful and time-consuming for users to discern meaningful labels for clustering outcomes 2) meaningful interpretations often depend upon how well users interact with clustering models. 
In addition, understanding complex ML output is mentally demanding, making it difficult for domain users (such as salespeople) to adopt and integrate such sense-making into their workflows~\cite{raees2023four, bae2020interactive}.
Making sense of data through interaction is beneficial, however, how effectively end users interact with such process is equally important to consider~\cite{mozolewski2022explain, raees2023four}.
Therefore, exploring interaction challenges for sense-making with agnostic clustering outputs is an important application area, specifically for non-technical domain expert users~\cite{pico2021bringing, raees2023four, moradi2022applications}.

The applications of customer segmentation are explored in several Business-to-consumer-based (B2C) market research and sales domains~\cite{naim2023consumer, abbasimehr2022analytical, manjunath2021distributed}.
We anticipate that similar methods are adapted to B2B segmentation contexts to address challenges and explore strategies to integrate segmentation effectively within organizational frameworks~\cite{cortez2021b2b}. 
The segmentation approaches explore extracting complex data patterns to inform decisions, however, effectively integrating B2B segmentation with domain (sales) users is tedious~\cite{pico2021bringing, moradi2022applications}.
For instance, B2B system integration for domain users in the context of the human-ML paradigm remains to be explored further to alleviate domain users' interaction with ML models~\cite{raees2023four, moradi2022applications}. 
Such interaction relates to designing ML tools that are interpretable and useful for non-technical users in business contexts, using iterative design and evaluation processes to meet their needs effectively.

Hence, this case study explores integrating unsupervised learning into the B2B customer segmentation context, enabling sales experts to interactively label and adapt ML outputs.
We leverage realistic datasets, often inaccessible to academics, to combine semantically-agnostic clustering algorithms with subjective domain expertise for ML-assisted decision support.
We conducted the case study evaluation in two phases.
In the first phase, we assess domain users' needs and organizational factors that influence transitioning to ML-based segmentation. 
We designed an interactive prototype to elicit user requirements while they experiment with customer segmentation using unsupervised ML algorithms.
We experimented with feature engineering, clustering methods, and interaction interfaces with sales (domain) users.
Afterward, we enhanced the prototype to allow salespeople to create their customer segmentation by choosing features and then visually inspecting the created clusters. 
In that development process, we identified the need for end users to label clusters with semantic understanding. 
Overall, this case study presents both technical prototypes, developed to identify domain users' needs and then to cater to those needs.
Our research advocates for user-centered HCI approaches in business while designing ML applications to match end-users context in B2B manufacturing workflows.
Throughout this process, we report learned lessons and insights while deploying this IML prototype in the sales customer segmentation context.

\section{Background}
The growth of ML adoption has increased across business contexts, for instance in areas such as analytics, sales, and forecasting~\cite{moradi2022applications, xu2022leveraging}. 
ML techniques---both supervised and unsupervised---are used in businesses to get insights that improve processes such as advertising, profiling, marketing, etc.~\cite{shen2021commerce, naim2023consumer, schlogl2019artificial}. 
However, integrating ML implementations with domain users presents challenges related to interaction, technical understanding, and organizational dynamics~\cite{moradi2022applications, iyelolu2024implementing, smith2024identifying}.
Business users often lack complex ML understanding and rely on intuition and experience for decision-making~\cite{iyelolu2024implementing}.
For instance, in B2B environments, domain experts often limit themselves to subjective insights due to such complexity~\cite{raees2023four}.
Hence, in a business context, addressing technical and acceptance limitations for smooth ML integration with domain users is necessary~\cite{iyelolu2024implementing}.
Such integration can support users to interpret and adapt ML models, for instance, to improve mapping between anticipated and actual results and explore insights with expertise.
Understanding user behavior within contextual B2B segmentation domains, where subjective insights are integral, is also important.

Agency over ML systems can help adjust models to domain (business) experts' needs~\cite{moradi2022applications, raees2023four}.
However, to make adjustments, systems must support interaction with models according to user expertise~\cite{raees2024explainable, qian2018grounding}.
This way, users can be empowered to interact with ML effectively and align it to their preferences.
For instance, interactive visual cluster explorations can be used to augment models with domain expertise~\cite{bae2020interactive}.
In unsupervised models, the iterative sense-making process can benefit from the subjective interpretation through interaction~\cite{bobek2023knac, iyelolu2024implementing, mozolewski2022explain}.
In addition, understanding clustering results through visualizations is also effective for sense-making~\cite{raees2023four, bae2020interactive}.
These studies show the value of user interaction with unsupervised learning models to improve understanding and insights. 
However, while implementing such tools, it is also important to carefully evaluate diverse domain perspectives to build effective interaction to support smooth integration. 

From a business perspective, segmentation methods categorize data based on defined attributes~\cite{manjunath2021distributed}, applying statistical methods (e.g., K-Means~\cite{zhao2021extended}, DBSCAN~\cite{xu2015comprehensive}, hierarchical clustering~\cite{manjunath2021distributed}) to group customers into different segments. 
However, many developments require adaptation to market needs or expert intuition, limiting their effectiveness in dynamic contexts. 
Adaptation to dynamic needs can help capture nuances and requirements for domain experts. 
For instance, B2B client relationships often involve long-term commitments (with manufacturers) that sales experts sometimes infer from client interactions but do not capture them in the (quantitative) data.
Considering the factors involved such as long-term relations and dynamic needs, B2B context contexts also require models to be adaptive~\cite{moradi2022applications}.

In interactive ML segmentation, users can adapt and correct models with direct modifications using their expertise~\cite{raees2023four, mozolewski2022explain} that allows them to build confidence and flexibility to create customized outcomes.
For instance, Gao et al.~\cite{gao2024evaluating} evaluate interactive topic segmentation with domain experts, highlighting that interactivity supports user understanding.
Their work~\cite{gao2024evaluating} explains that automated approaches for segmentation may not fully capture the model's real-world benefits without domain users' validation. 
Some studies have explored the impact of interactive segmentation on business contexts involving domain users~\cite{raees2023four, pico2021bringing, iyelolu2024implementing}.
For example, research by Pico-Valencia~\cite{pico2021bringing} indicates that interactivity empowers non-technical business users with ML models through Graphical User Interface-based (GUI) interactions.
Similar interactive solutions can help alleviate the issues related to understanding statistical models and adapting them with user insights, for instance, in domains like B2B customer segmentation~\cite{bae2020interactive, raees2023four}.

In summary, studies~\cite{bobek2023knac, gao2024evaluating, iyelolu2024implementing} explore the potential of designing effective interaction for domain experts ML segmentation.
The interactive segmentation integrates user expertise with ML, promoting a human-in-the-loop ML approach that supports iterative feedback, minimizing gaps between model outputs and user expectations.
While automated ML approaches are explored, customer segmentation in B2B contexts necessitates human interpretation and agency~\cite{gao2024evaluating, raees2023four}, due to involved complex expertise-dependent decision-making. 
The knowledge and interaction gap between domain users and systems (models) also limits effective decision-making support.
Interpretable and interactive ML tools can also benefit from user interactions to improve their performance through an iterative feedback loop.
Hence, recognizing where and how domain experts can provide value is critical to improving ML-based decision support.

\section{Example Scenario}
\label{app:scnrio}
Consider Alice, a sales analyst, who needs to perform customer segmentation to analyze company customers, for instance, to prioritize orders, services, or reporting for strategic planning.
She finds that existing rule-based and subjective solutions are not sufficient to cater to her objectives. 
Alice utilizes an unsupervised ML segmentation model to perform analysis, understand the segments, and make informed decisions.
However, to interpret the output from the resultant clusters she must carefully analyze the segments using features, make associations, and then draw conclusions to make the analysis useful.
She has domain expertise and subjective knowledge beyond features to discern individual customers. 
Therefore, with agnostic unsupervised learning models (being limited in assigning meaningful labels), she has to experiment with model outputs for better understanding.
Without semantic links or detailed (laborious) exploration, she finds it difficult to label segments and build objective rationale meaningfully. 
Also, if the model is not adaptable, she cannot adjust it to her context to include market factors, segmentation variables, or models. 
Therefore, Alice would benefit from a custom and flexible approach that can provide her with options such as: 
\begin{itemize}
    \item Customizing the segmentation solution and lowering the barrier for her to adapt to dynamic needs by graphically adjusting parameters and interface elements. 
    \item Semantically label or correct the clustering outcomes to make them meaningful by defining the feature to label associations. Alice can rationalize and make sense of adaptations without being constrained by fixed/static ML output.
    \item With repeated trials, Alice can compare and contrast segmentation versions strengthening her analysis for a more robust understanding of the data. Alice can improve and complement her sense-making process with iterative evaluation. 
\end{itemize} 

\section{Eliciting Requirements for an Interactive ML Customer Segmentation Tool}
Our context is Sappi, a global company focused on dissolving pulp, paper pulp, and bio-refinery solutions to its customers in over 150 countries, having teams in multiple regions.
To understand sales experts' practices in terms of how customer segmentation is used in their work, we followed a technology probe approach~\cite{hutchinson2003technology}. 
We designed a prototype for ML clustering and conducted expert sessions with participants about how they perceive, understand, and want to adapt such models to their contexts. 
We also wanted to map existing practices to reduce users' effort and learning curve. 
Hence, we set a broad objective to understand domain users' insights in creating a segmentation model and utilizing it to make decisions for their daily work.

\subsection{Prototype Development}
\begin{figure*}
  \centering
  \includegraphics[width=\textwidth]{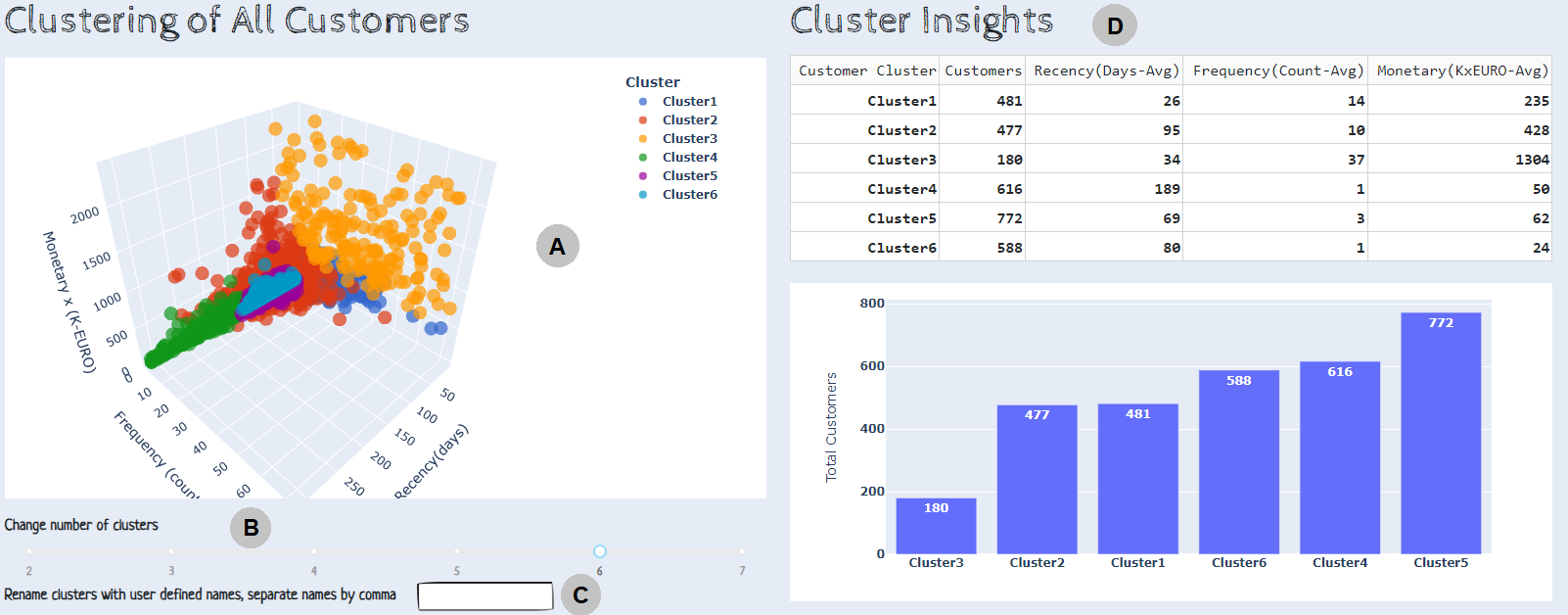}
  \caption{Interactive cluster analysis allowed users to A) visualize, interact, and filter with the graph (and its elements) to make sense of data by tweaking visualization elements, B) select models (by changing cluster counts) to contextualize output, C) Modify or re-label segments to make connections and contextual understanding, D) visualize and an overview of descriptive statistics about each cluster in the selected model to overall understanding. For confidentiality reasons, the visualization is modified with anonymous data and labels.}
  \Description{Clustering visualization shows how the prototype can be used by changing the clusters, visual interaction, and filters. You can visualize the segmentation in a 3-D graph. The figure shows a 3D visualization and controls to adapt it.}
  \label{fig:case1}
\end{figure*}

To create the customer segmentation we worked with a real-world dataset from a sales context.
The dataset contained around ten million business transactions (over the last four years), having temporal and categorical features.
The data was batched into sub-markets based on regions to make personalized models. 
We experimented with various clustering methods (K-Means, DBSCAN, hierarchical~\cite{zhao2021extended, xu2015comprehensive, manjunath2021distributed}), and selected K-Means and DBSCAN~\cite{zhao2021extended, xu2015comprehensive} combined with RFM~\cite{hughes1994strategic} segmentation model for further evaluations. 
For simplicity, we adapted an RFM segmentation (a commonly used model) to B2B contexts~\cite{abbasimehr2022analytical}.
We created interfaces to support users for interaction and embedded them in a web-based application using React~\cite{react2024} and Plotly Dash~\cite{dash2024}. 
The prototype, accessed through a browser, allowed interaction between components through the Dash callback functions~\cite{dash2024}.
Figure \ref{fig:case1} shows an example interface with a segmentation model and its visualizations, allowing users to reflect upon the clustering results.
The interaction allows users to adjust filters, drill down segmentation levels, change clusters, and relabel them. 
We trained models on datasets for a range of cluster settings (counts, labels, etc.) and allowed users to experiment with all of those.
Cluster settings such as input data and models were pre-defined, however, users could control the cluster count, control visualizations, and modify labels to improve their sense-making and analytics. 

\subsection{Participants and Procedures}
We recruited seven domain experts (\textit{5 males, 2 females}) having years of experience (\textit{mean=18.5, std=7.9}) in various leading roles in the company (as shown in Table \ref{tab:users1}).
The participation was voluntary and the process followed the Institutional Review Board (IRB) guidelines and company policies on data use and consent. 
During experiments, we provided experts with a demo of how the prototype works and asked them to explore it by themselves. 
We conducted in-depth interviews (\textit{mean=53.5 minutes, std=13.8 minutes}) with participants to explore 1) how they use such analytics in their work contexts; and 2) the impact of their decisions based on such customer segmentation.
These experiments were conducted through online interviews with the prototype exploration. 
We provided sample tasks to participants, for instance, to use segmentation models, adjust those to their needs, connect to their existing work, and make their own decisions based on system outputs.
In addition, we asked participants to report their experiences with a post-study questionnaire.
\begin{table*}
  \caption{Participants' demographics having diverse expertise in the company and sales domain.}
  \label{tab:users1}
  \Description{This table presents the demographics of participants with diverse expertise in the company and sales domain. It includes columns for participant ID, gender, title, region, analytics tools used, and years of experience.}
  \begin{tabular}{llllll}
    \toprule
    ID&Gender&Title&Region&Analytics Tools&Experience (Years) \\
    \midrule
    FP1&M&Market Manager&North America&Salesforce, SAP&15 \\
    FP2&M&Sales Manager&Europe&Tableau, BOA&30 \\
    FP3&F&Sales Manager&Europe&Salesforce, Excel&10 \\
    FP4&F&Head of Sales&Europe&Salesforce, Excel&20 \\
    FP5&M&Head of Sales&Europe&Salesforce&10 \\
    FP6&M&VP Sales&North America&Salesforce&30 \\
    FP7&M&Marketing Director&North America&Tableau, SAP&15 \\
    \bottomrule
\end{tabular}
\end{table*}

\subsection{Results}
Using a mixed-methods approach we collected data using a questionnaire (5-point Likert scales) and interviews.  
We analyzed survey responses and transcribed interviews for feedback analysis.
We asked participants about their self-reported experiences with the prototype and how ML assistance can support them. 
Most participants showed positive impressions of prototype adoption for insight creation and segmentation.
In general, participants valued ML segmentation for decision support (\textit{mean=4.0, std=0.53}) and the perceived quality of derived data (\textit{mean=3.42, std=0.49}) for strategic and tactical operations. 
However, participants indicated the need for adaptation of ML-based segmentation to derive valuable insights in similar scenarios (\textit{mean=3.85, std=0.35}).

These insights provided an overall user impression of interactive exploration, however, we emphasized more on the qualitative analysis as we needed to extract a deep perspective on the participants' experiences. 
We thematically analyzed participant interviews to identify patterns and grouped them to collect observations (e.g., prototype explanations).
The analysis revealed that participants' agency over the prototype to map with user input is necessary, and it was important that the final decision-making resides with users.  
Participants' quotes in the following paragraphs explain the findings from interview sessions. 

Participants expressed on making the prototype more expressive and including refined control to rationalize decision-making, such as using explanations, making corrections, validating insights, etc. 
For instance, if they had to explain how a decision is made based on ML outcomes, they should be able to support that with reasoning or logic.
This was particularly important to rationalize their existing/background understanding of specific data examples (customers). 
Importantly, domain users wanted to make models more interactive and adaptive in execution and sense-making.
FP1 explained, \textit{“Promising (segment) does not make sense to me here ... I would have a better definition around it and would change it”}.
In addition, FP6 highlighted, \textit{“To me, I would interact with labels and clusters, to find someone (a customer) that I can target to sell ... labels are just the starting point to shape the analysis”}.
Particularly, the cluster labeling option triggered conversations to improve and make it more adaptable, for instance, to build some connections with the labels. 
Users visually interacted with models and provided feedback on their understanding and rationale. 
For instance, sales experts cross-checked their subjective impressions about valuable customers. 
FP5 explained it as \textit{"I could use labeling to follow the tail ... that is the customers who might be interested if offered better value"}.
FP1 also highlighted this as: \textit{“If I have some products left to sell towards the end; I have to look for best 10 in a segment”}.
This practice helped us start identifying requirements for interactive customer segmentation. 
Varying insights from different participants helped us infer the dynamic needs for exploratory development.  

\subsection{Reflections}
We realized that some features (e.g., recency) were less important for strategic customers. 
To many experts, feature variety could contribute to defining clusters, and not all of them have to be used statically.
Therefore, customization is important and should be adaptive at the run time, depending upon the scenario and market situation. 
For instance, some customers provide good revenue but have higher costs to serve, so they might not be the top priority. 
We identified that there were slight nuances in definitions and standards for customer segmentation across participants. 
For instance, in low-performing markets and products, segmentation needs to be adapted to maintain a presence, further highlighting the need for interactive adaptations. 
Sales and data-driven insights can be combined to override model outcomes, but ensuring the validity of outcomes is challenging without domain expertise. 
A key criterion for integrating domain expertise with models is to first explore how customers are given a label, for example, by identifying features that categorize a group of customers into a particular segment.
Even with the diversity in B2B sales participants' markets and objectives, the common theme revolved around approaches for labeling clusters with their contextual relevance.

\section{Piloting an Interactive ML Customer Segmentation Tool}
With the analysis of the first deployment, we re-evaluated requirements and re-designed the prototype, improving model interactions, explanations, usability, and other information visualization features. 
We designed a semantic labeling process that allows users to rationalize segment labels with features. 

\subsection{Prototype Enhancement}
\begin{figure*}
  \centering  
  \includegraphics[width=\textwidth]{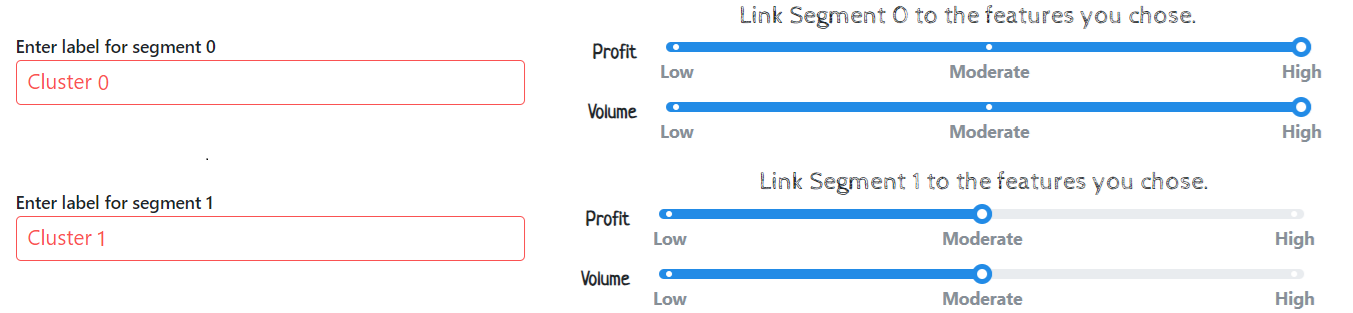}
  \caption{Users can type and map their labels with the model’s features -in this case, profit and volume of products bought. For instance, in this case, the user specifies a high profit and volume of products bought for cluster 0 (i.e., one can imagine this cluster grouping really good customers) and a moderate profit and volume of products bought for cluster 1 (i.e., one can imagine this cluster grouping customers who might have the potential to grow in terms of the profit and volume of products that they buy.)}
  \Description{The process of label mapping uses a text-to-rating scale where each label can be given a mapping with every selected feature to specify its importance.}
  \label{fig:case2_1}
\end{figure*}

We worked further with the dataset and included other temporal and categorical variables so that our end users could have a larger set of available features to choose from for their customer segmentation. 
These features included profit, volume (of sold products), product group, and inter-purchase interval, among other features.
We developed the pipeline allowing users to build the ML model by selecting input data, for example, filtering the timeframe for which they wanted to create the segmentation, and allowing them to name labels themselves and map those labels to the features they chose before the model is created by the algorithm, as shown in figure \ref{fig:case2_1}.
This helped users rationalize feature importance to segmentation, which otherwise, would require significant effort to make sense. 

\begin{figure*}
  \centering
  \includegraphics[width=\linewidth]{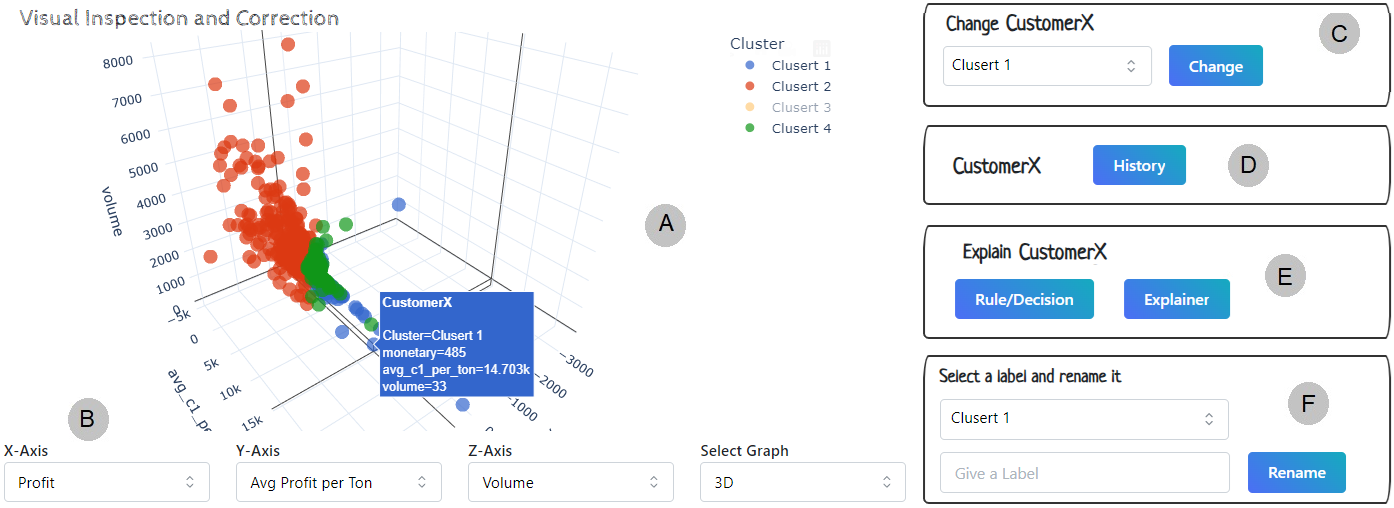}
  \caption{Inspection allows users to modify and adapt segments and use features to support their decision-making. Users can A) interact with and adjust the visualization to inspect the segmentation from different angles B) visualize several graphs and modify visualization by selecting the features they want to display C) modify the selected instance from the visualization D) visualize the historical information of the selected instance E) view various explanation visualizations to inspect models F) Relabel a cluster. For instance, a user is visualizing "CustomerX" by adjusting visualization to 3D and selecting the features: profit, average profit per ton, and volume for each axis.  For confidentiality reasons, the visualization is modified with anonymous data and labels.}
  \Description{Visualization shows enhanced visual clustering, interaction, and filters. You can visualize the segmentation in a 3-D graph, view data examples, and change labels.}
  \label{fig:case2}
\end{figure*}

Once, the user specifies the semantic mapping, the system builds the segmentation, optimally linking feature significance to labels. 
Afterward, users can inspect, evaluate, or correct label mapping (as shown in figure \ref{fig:case2}) to further refine the results.
The inspection enhances existing interactive features with explanations, output correction, and complementary information to support decision-making.
We used rule-based decisions and a user-centered explanation model (LIME~\cite{ribeiro2016should}) to support user understanding. 
Users can also assess other information to support their analysis by viewing customer profiles and historical interactions and visualizing them using an adjustable scatter graph.
Model overriding provides individual and group correction features to adjust the label instances individually or collectively. 

\subsection{Participants and Procedures}
We recruited seven (not involved before) participants (\textit{6 males, 1 female}) with diverse domain and technical experience (\textit{mean=10, std=3.8}) as shown in Table \ref{tab:users2}.  
We followed consistent IRB protocols and company regulations for scenario experiments.
We asked participants to interact with the prototype following a systematic approach by 1) providing participants with a system overview, 2) giving them tasks to perform, and 3) capturing their experiences through an interview and a questionnaire.
The prototype allowed participants to build models interactively (e.g., to create clusters, and make semantic links, and corrections).
We asked participants to build segmentation models from scratch themselves using their datasets.
The process allowed users to select datasets, filter those based on their requirements, and then choose the features and number of segments they wanted to create. 
Then, users create the semantic links as shown in the figure \ref{fig:case2_1}, for each segment and select features before the system builds the segmentation. 
We asked them to reflect on system output with their understanding by utilizing information and visualization features.
Then, we asked them to evaluate the model they built using information and explanation features with their existing segmentation.
Interviews (\textit{mean=45 minutes, std=6 minutes}) with participants focused on assessing both the technical and domain functionality of the prototype.
The discussions and questionnaires focused on evaluating models and their acceptability while minimizing the adaptations required from users. 

\begin{table*}
  \caption{Details of sales (SE) and technical (TE) experts involved in the pilot sessions.}
  \label{tab:users2}
  \Description{This table presents the demographics of participants with diverse expertise for the second study. It includes columns for participant ID, gender, function, region, and years of experience.}
  \begin{tabular}{llllll}
    \toprule
    ID & Gender & Function & Region & Experience (Years) & Expertise \\
    \midrule
    SP1 & M & Head of Sales & Europe & 15 & Salesforce, Tableau, BOA \\
    SP2 & M & Sales Manager & South Africa & 15 & Salesforce \\
    SP3 & M & Sales Manager & Europe & 10 & Salesforce, Tableau \\
    SP4 & M & Marketing Head & North America & 10 & Tableau, BOA \\
    SP5 & M & Director Digital & Europe & 10 & ML, Project Management \\
    SP6 & M & Visual Analytics Lead & Europe & 5 & Analytics, ML \\
    SP7 & F & Analytics Lead & Europe & 5 & Tableau, Looker \\
    \bottomrule
  \end{tabular}
\end{table*}

\subsection{Results}
Again, we used the mixed-method approach to analyze the data collected through questionnaires (5-point Likert scales) and qualitative interviews. 
The analysis shows that interactive model building/inspection is highly effective in sales analytics. 
We evaluated the role of users' agency and explanations on the overall acceptance and quality of extracted information. 
Participants valued agency and support for interactive model building (\textit{mean=4.14, std=0.34}).
Rationalizing labels with features and the ability to make corrections were particularly useful for the participants, even after semantic mapping.
Likewise, participants showed that semantic mapping and explanations deepened their understanding (\textit{mean=4.28, std=0.45}). 
This indicates that interactivity, explanations, and semantic mapping help in sense-making and subsequently correct those labels irrespective of the initial setting.
Participants reported various strategies that can help them evaluate model outcomes.
For instance, several participants first labeled the whole segment and then explored individual instances where refinement was needed, particularly for boundary cases.  
The explanatory and complementary information helped them make fine-grain analyses for those specific cases in question.
For instance, interactive visualization is supported by drill-down features to examine a specific customer or compare it against others. 

Participants liked the way they could control and steer the model-building process for segmentation.
For instance, SP5 noted, \textit{"It is very fast to get out of what you build... Also, you can modify it as your market demands".} 
Participants highlighted that deeply exploring models gives them more confidence to experiment than one-time static deployments. 
As SP6 said, \textit{"It is understandable and it gives me the confidence to test out my hypothesis against different settings".} 
The evaluation feature also supports users in enhancing their confidence in the system by inspecting explanations for each case. 
SP7 explained it as, \textit{"For any assumptions, I could test it by benchmarking feature thresholds, I would demote a customer to a lower rank not meeting those".} 
Semantically linking cluster names to feature dominance provided clarity and rationale to participants.
Pre-conceived knowledge of participants about segmentation and features also dictated their experiences. 
For instance, creating a semantic mapping provides them with knowledge and expectations of desired clusters in advance. 
For example, SP5 mentioned, \textit{"Higher profits should have moved one segment separate from the others, I think other features have impacted higher too"}.

\section{Discussion}
We experimentally observe that customer segmentation is a complex and high-stakes domain, requiring experts to steer it for optimal decisions.
Hence, participants needed flexibility and subjective adjustment to build deep rationale.
For instance, experts must reason their understanding of each decision (assigning a label to a customer) to substantiate its effectiveness. 
We evaluated the effectiveness of interactivity through our prototype implementation and extracted directions for further improvement. 
Our results show the value of interactive semantic labeling for reducing the effort required to understand cluster outcomes, however, we anticipate making the assessment more robust.
We also report the experiences and nuances of domain (sales) users in their interactions with ML segmentation.
Participants highlighted further requirements to improve the prototype in several contexts.
For instance, we engineered several features from datasets, however, salespeople often have subjective insights about some data (really good or bad customers) and exclude them from the modeling process.
This requires additional filtering in the data pipeline for analyzing outliers before building models. 
In addition, participants liked to have options to curate new features during runtime which would further enhance control and adaptations to the context. 
Participants highlighted that the prototype should be more assistive in terms of highlighting incorrect outcomes by default to speed up the sense-making and correction process.
However, with unsupervised models (without ground truth), it is hard to validate the correctness of individual cases. 
Yet, we acknowledge and aspire to improve features to enhance user experience and ease their transition to ML segmentation. 

While this work focused on B2B customer segmentation in a global manufacturing context, our findings can be generalized to other domains where users face similar problems with ML outputs. 
The need for interactive semantic mapping and the value of iterative model refinement is applicable across business scenarios such as professional services, healthcare, or high-tech industries. 
Moreover, the challenges we encountered in transitioning domain experts to ML-based decision support systems are also common across industries adopting ML technologies. 
However, it's important to note that the specific features and interaction patterns may need adjustment for different use cases.

\subsection{Lessons}
We evaluated users transitioning to ML-based customer segmentation using unsupervised methods that require continuous subjective adjustments. 
First, we evaluated decision-making automation for customer segmentation, however, the domain complexities in B2B segmentation make the automation case unsuitable. 
In our context, decision-making included varied definitions and layered abstractions of subjective expertise throughout the organization.
Therefore, we resort to a more interactive implementation, which is crucial in contextual domains such as B2B sales analytics where domain knowledge plays a significant role in decision augmentation.
Based on users' requirements, we evaluated the dynamic ML segmentation assistant, allowing them control over the ML process, however, concerns remain for effective integration.  
For instance, semantic mapping with users' (subjective) vocabulary is important, which most unsupervised approaches are susceptible to.
However, in scenarios, where data is unlabeled and features change over time, unsupervised methods (e.g., clustering) are effective in making sense of data. 

We realized that only a data-driven approach was not enough for salespeople to make decisions.
Other factors influencing decisions such as customer relationship management, which is an important area to provide insights to salespeople beyond models that must be accorded through interaction. 
In general, users in such settings find it hard when algorithmic results are not adjustable or map with their intent.
Hence, agency and effective interaction are necessary in such contexts to adjust models to users' intent.
 
In some cases, we found that overly depending on subjective knowledge increases the likelihood of personal biases.
Users' self-efficacy and background knowledge can overshadow the results generated by the model which subsequently can decrease the user reliance on the model.
Therefore, balancing subjective and objective aspects is necessary, for instance, by reducing the gap between users and algorithms with iterative statistical validations and semantic mapping to facilitate algorithm adoption.
During the development, we understood that it would be very hard to label segments appropriately without deep expertise in the domain and having limited observability of existing semantic mapping. 
Hence, enabling domain users to steer ML models with a feedback loop can help achieve their goals.

Experimenting with actual users initially in the design process was beneficial and it informed adding features not envisioned initially.
With such analysis, we identified the usability and explanation needs of users having low-ML expertise. 
Therefore, we used various assistive cards, visual cues, and tooltips to support quick-help functionality.
Such intuitive interface features help improve information understanding and usability for end users.
It is also essential to select appropriate contextual settings as general use cases with open-source data may not generalize well with complex user (business) domains.
Had we selected a general use case, it would not have made contextual sense with end users and their workflows. 
Therefore, it is good practice to perform a phenomenological analysis with the target users in a work environment to better map the context. 
 
\subsection{Limitations and Future Work} 
We made iterative efforts to reduce usability issues, however, there is still a need to refine visual elements for better understanding. 
We aspire to build more intuitive designs in the future to better connect with the user base. 
For instance, we observe that although interactivity supports flexibility it also burdens the user to take action. 
We evaluated the labeling outcomes on users' subjective and qualitative measures reporting better satisfaction.
Although this supports initial evidence for our method, we aim to conduct a more robust statistical evaluation.
We also foresee that domain experts prioritize customers based on the assumption that they want to grow.
The idea of quantifying such soft factors is difficult to manage with objective models, hence, we accommodate that through interactions for now. 
Future work can include exploring measures to quantify these soft factors and integrate them with models. 
Enhancing models to include nominal variables is also an important direction. 
We will explore the applicability of our approach in other segmentation contexts (e.g. for segmenting patients in the medical domain, or learners in education) to further validate and improve generalizations.
We had a few participants with diverse expertise, however, we anticipate increasing the sample size and enhancing validation methods (both clustering measures and subjectivity) to improve applicability.

\section{Conclusion}
Domain users oftentimes face challenges with technical implications to interpret the clusters produced by an unsupervised Machine Learning (ML) algorithm. 
We identified domain users' requirements to understand their current practices and developed an interactive prototype to support their transition to ML-based customer segmentation.
We evaluated the effectiveness of semantically mapping labels to features to build understanding and rationalize cluster outputs for further inspection. 
Our solution enables sales users to build segmentation models with enhanced agency and sensemaking with unsupervised ML models.
Future work includes enhancing the prototype to cater to diverse requirements and integrating aspects of customer relationships by quantifying subjective and qualitative user experiences.
Our work informs lessons for IML systems in customer segmentation and perhaps other related contexts to stimulate additional research endeavors. 

\begin{acks}
    This research has been sponsored by Sappi in the context of the Sappi-RIT Digital Innovation Lab at Golisano College of Computing and Information Sciences, RIT. Any explanations, findings, or conclusions expressed in this work are those of the authors and do not necessarily reflect the views of our sponsor. We would like to extend our gratitude to our Sappi colleagues Markie Janse van Rensburg, Kouris Kalligas, Pieterjan Geens, Kamila Grabowska, and Nicky Bachert for their continued support in the work reported in this paper.
\end{acks}

\bibliographystyle{ACM-Reference-Format}
\bibliography{sample-base}

\appendix

\end{document}